\documentclass[oldversion]{aa}     % use v6.0 with [oldversion] or aa_urb.cls without
\def \th {\thinspace}

\def\approxgt{\mathrel{\hbox{\rlap{\lower.55ex \hbox {$\sim$}} \kern-.3em \raise.4ex \hbox{$>$}}}}
\def\lesssim{\mathrel{\hbox{\rlap{\lower.55ex \hbox {$\sim$}} \kern-.3em \raise.4ex \hbox{$<$}}}}
\def\approxlt{\mathrel{\hbox{\rlap{\lower.55ex \hbox {$\sim$}} \kern-.3em \raise.4ex \hbox{$<$}}}}
\def \degmark {^\circ}
\def \sun {\hbox {$\odot$}}
\usepackage{epsf}
\usepackage{graphicx}
\begin{document}

\title{Dipping in Cygnus\th X-2 in a multi-wavelength campaign 
          due to absorption of extended ADC emission}

\author{M. Ba\l uci\'nska-Church\inst{1}
\and N. S. Schulz\inst{2} 
\and J. Wilms\inst{3}
\and A. Gibiec\inst{4}
\and M. Hanke\inst{3}
\and R. E. Spencer\inst{5}
\and A. Rushton\inst{5,6,7}
\and M. J. Church\inst{1}}
\institute{
           School of Physics \& Astronomy, University of Birmingham,
           Birmingham, B15\th 2TT, UK\\
\and      
          Kavli Institute for Astrophysics and Space Research, Massachusetts Institute of Technology,\\
          Cambridge MA\th 02139\\
\and
          Dr. Karl Remeis-Sternwarte, Astronomisches Institut der Universit\"at Erlangen-N\"urnberg,\\
          Sternwartestrasse 7, 96049 Bamberg, Germany\\
\and
          Astronomical Observatory, Jagiellonian University,
          ul. Orla 171, 30-244 Cracow, Poland\\
\and      
          Jodrell Bank Centre for Astrophysics, School of Physics \& Astronomy,\\
          University of Manchester, Manchester M13\th 9PL\\
\and      
          Onsala Space Observatory, SE-439 92 Onsala, Sweden\\
\and
          European Southern Observatory, Karl-Schwarzschild-Strasse 2, 85748 Garching, Germany\\}
\offprints{mbc@star.sr.bham.ac.uk}
%\thanks{}

\date{Received 14 October 2010; Accepted 22 March 2011}
\titlerunning{Cygnus X-2 with XMM and Chandra}
\authorrunning{Ba\l uci\'nska-Church et al.}

\abstract{We report results of one-day simultaneous multiwavelength observations of Cygnus\th X-2 using
{\it XMM}, {\it Chandra}, the European VLBI Network and the {\it XMM} Optical Monitor. During the
observations, the source did not exhibit Z-track movement, but remained in the vicinity of the soft apex.
It was in a radio quiescent/quiet state of $<$ 150 $\mu $Jy. Strong dip events were seen as 25\%
reductions in X-ray intensity. The use of broadband CCD spectra in combination with narrow-band 
grating spectra has now demonstrated for the first time that these dipping events in Cygnus\th X-2 are 
caused by absorption in cool material in quite a unique way. In the band 0.2 - 10 keV, dipping appears to be due
to progressive covering of the Comptonized emission of an extended accretion disk corona,
the covering factor rising to 40\% in deep dipping with an associated column density of $3\times 10^{23}$
atom cm$^{-2}$. Remarkably, the blackbody emission of the neutron star is not affected by these dips, in 
strong contrast with observations of typical low mass X-ray binary dipping sources. The {\it Chandra} 
and {\it XMM} gratings directly measure the optical depths in absorption edges such as Ne K, Fe L, and 
O K and a comparison of the optical depths in the edges of non-dip and dip data reveals no increase of 
optical depth during dipping even though the continuum emission sharply decreases. Based on these findings, 
at orbital phase 0.35, we propose that dipping in this observation is caused by absorption in the outer disk by structures 
located opposite to the impact bulge of the accretion stream. With an inclination angle $> 60^{\degmark}$
these structures can still cover large parts of the extended ADC, without absorbing emission from the 
central neutral star.
%{\bf and on the orbital phase of $\sim$0.35, we propose that the dipping in this observation is caused by 
%the neutron star, so that this is not absorbed, also demonstrating that the ADC cannot be point-like.}
}
% keywords must be inside abstract
\maketitle

\section{Introduction}

Z-sources form the brightest group of low mass X-ray binaries (LMXB) containing accreting
neutron stars with super-Eddington luminosities between $2 - 6 \times 10^{38}$ 
erg s$^{-1}$. Strong physical changes take place within the sources as is
shown by the three spectral tracks traced in X-ray hardness-intensity diagrams known as the 
horizontal branch (HB), the normal branch (NB) and flaring branch (FB) (Schulz et al. 1989; 
Hasinger \& van der Klis 1989). The six original Galactic Z-sources are divided into the Cygnus\th X-2 
like sub-group (Cygnus\th X-2, GX\th 340+0 and GX\th 5-1), in which complete Z-shapes 
are normally seen, and the Sco\th X-1 like sub-group (Scorpius\th X-1, GX\th 17+2 and 
GX\th 349+9) in which the HB is small or missing but in which flaring is stronger (Hasinger \& 
van der Klis 1989; Kuulkers \& van der Klis 1995). It is important to understand the formation of 
relativistic jets in these sources. In Sco\th X-1, radio observations with the VLA 
revealed the outward motion of radio condensations at v $\sim$0.45c (Fomalont et al. 2001). 
The radio detection is on a restricted part of the Z-track near the hard apex between HB and NB. 
When X-ray bright, the highly variable LMXB Cir\th X-1 exhibited Z-source behaviour (Shirey et al. 1997), 
a powerful X-ray accretion disk wind (Brandt \& Schulz 2000), and relativistic jet emission 
(Fender et al. 2004, Heinz et al. 2008). Thus X-ray observations of Z-sources uniquely
allow the study of conditions connecting the inner disk and the neutron star to mechanisms of 
jet formation.

The nature of the Z-track phenomenon is not understood. However, recent studies have 
produced models that can explain spectral changes along the Z-track, notably the
extended accretion disk corona (ADC) emission model of Church \& Ba\l uci\'nska-Church (2004).
This is based on long-term studies of the dipping LMXBs in which absorption in the bulge 
in the outer disk causes dips in X-ray intensity at the orbital cycle (White \& Swank 1988). 
Spectral evolution during dipping can be explained in terms of point-like blackbody emission 
from the neutron star and the dominant Comptonized emission of an extended ADC (Church \& 
Ba\l uci\'nska-Church 1997). The extended nature of the ADC is shown by measurement of dip ingress 
times giving sizes of 20\th 000 - 700\th 000 km depending on luminosity (Church \& 
Ba\l uci\'nska-Church 2004). Strong support for an extended ADC comes from {\it Chandra} 
high-resolution grating results (Schulz et al. 2009) for highly excited H-like lines
of Ne, Mg, Si, S and Fe in Cygnus\th X-2, with Doppler widths corresponding to 
radii between 18\th 000 and 110\th 000 km in good agreement with dip ingress timing. 

The extended ADC model produces a straightforward explanation of the Z-track
in the Cyg\th X-2 like sources (Ba\l uci\'nska-Church et al. 2010). The large increase in
the luminosity of the ADC emission ($L_{\rm ADC}$) suggests that the mass 
accretion rate ($\dot M$) increases between soft and hard apex, opposite to previous 
suggestions (Hasinger et al. 1990) in which $\dot M$ decreases. The 
neutron star flux increases from a fraction of Eddington at the soft apex to 
super-Eddington at the hard apex and HB. The resulting high radiation pressure 
at the inner disk diverts accretion flow vertically launching the jets observed
on this part of the Z-track. The FB consists of unstable nuclear burning of the accretion flow 
as suggested 
by the good agreement of the mass accretion rate per unit area with the critical  
$\dot m_{\rm ST}$ when burning becomes unstable (Bildsten 1998).

\emph{RXTE} observations of a huge outburst in XTE J1701-462 (Lin et al. 2009) show the source 
evolving from a super-Eddington Cyg\th X-2 like source to Sco\th X-1 like, and finally into an 
Atoll source. The interpretation was that $\dot M$ changes caused the luminosity decrease, but 
did not drive Z-track motion. However, the observations did not produce much interpretation for the nature
of the Z-track, and in the spectral model used the Comptonized emission was only 10\% of the total
luminosity contrary to general acceptance that this is much higher in LMXB sources.

Cygnus\th X-2 is the archetypal Z-source. For a distance of 8 - 11 kpc (Cowley et al. 1979; Smale 1998)
it is thought that there is an evolved companion of 0.4 - 0.7 $M_{\sun}$. Recent high-resolution optical 
spectroscopy of Casares et al. (2010) provides a refined orbital solution with a period of 
\hbox{9.84450$\pm$0.00019 d.} Assuming an inclination of 
62.5$\pm$0.4 $\degmark$ from ellipsoidal fits to UBV lightcurves (Orosz \& Kuulkers 1999), Casares et al.
found a neutron star mass of 1.71$\pm$0.21 $M_{\sun}$, supporting a previous result (Casares et al. 1998),
while Elebert et al. (2009) derived 1.5$\pm$0.3 $M_{\sun}$.

Many observations of Cyg X-2 indicate an extended ADC. Vrtilek et al. (1988) suggested that
high and low flux states reflect changes in the geometrical and optical thickness of the accretion 
disk and ADC. H-like Fe~{\sc xxvi~} line emission was found by Smale et al. (1993, 1994)
using \emph{BBXRT}; lines from various K-shell ions were found (Kuulkers et al. (1997, di Salvo et al. 2002)
using \emph{ASCA} and \emph{BeppoSAX}.
Recent \emph{Chandra} observations revealed a wealth of detail: K-shell emission with lines from
H-like and He-like ions of Mg to Fe (Schulz et al. 2009).
%Mg~{\sc xii~} to Fe~{\sc xxvi~} (Schulz et al. 2009).
The fully-resolved broad lines gave Doppler velocities indicating 
%of 1100 - 2700 km s$^{-1}$
radii of more than 10$^9$ cm (as detailed above) and plasma densities of $\sim 10^{15}$ cm$^{-3}$
consistent with an origin in a hot, dense, extended ADC.
% of $T$ $>$ $10^6$ K.
%Furthermore, line fluxes varied along the Z-pattern indicating heating 
%along the branches with heating luminosities changing from 0.4 to 1.4 L$_{Edd}$. 
%However, all these observations also detected a very broad line feature at 971 eV,
%which is also present in the \emph{Chandra} data as an unresolved
%line complex. While Smale et al. 1994 suggested an Fe L line complex, its
%true origin remains unknown.

Cygnus\th X-2 often exhibits short-term decreases (dips) in X-ray intensity generally 
described as a feature of the FB unrelated to absorption, first recognized 
by Bonnet-Bidaud \& van der Klis (1982). 
%Vrtilek et al. (1988) found dips lasting from minutes 
%to hours. 
Hasinger et al. (1990) showed the reduced intensity intervals in two forms.
In hardness-intensity, the FB is usually short or absent, while
dipping comprises reductions in intensity from positions near the soft apex. But
in hard colour {\it versus} soft colour, the same dip data form a third track looking exactly like 
an extensive FB. Thus, there was confusion whether dipping was absorption or flaring, and showing 
the use of colour-colour diagrams may mislead because these do not distinguish between
decreasing and increasing intensity. Clarifying this has not previously been carried out. 
%Kuulkers et al. (1996) described the phenomenon as plain intensity reductions. 
The assumed association of dipping and flaring led to
the idea that inclination was the main difference between Cyg-like and Sco-like sources:
it was argued that in Cyg\th X-2 and GX\th 340+0, the FB corresponds to X-ray dips and that the 
Cyg-like group have high inclination allowing absorption in the inner disk inflated by radiation 
pressure (Kuulkers \& van der Klis 1995); in the Sco-like sources the inclination is small. 
This assumes that $\dot M$ is maximum on the FB causing disk inflation, but 
Ba\l uci\'nska-Church et al. (2010) proposed that $\dot M$ is minimal on the FB in the Cyg-like sources. 

There has since been little discussion of the nature of the dipping/flaring.
In the present work, we resolve this, showing that dipping is not associated with the FB. 
We present results of long simultaneous observations of Cyg\th X-2 with {\it XMM}
and {\it Chandra} aimed at investigating changes of the continuum and lines around the Z-track.
Simultaneous high-resolution radio observations were conducted, aimed at detecting relativistic jets,
if present during our campaign, and possibly monitor the ejection of plasma blobs from the source.

\section{Observations and data analysis}

We observed Cygnus\th X-2 in a multi-wavelength campaign on
2009 May 12 -- 13, starting at 9:30 UT. Both {\it XMM-Newton} 
and {\it Chandra} were used in essentially simultaneous one-day observations. In addition,
the Optical Monitor onboard {\it XMM} provided optical data in the UVW1 band, i.e. in the band  
2500 - 4000 \AA.
 
High-resolution interferometric radio data were obtained using the
e-VLBI mode of the European VLBI network (e-EVN) at 5~GHz. Observations
were approved following a target-of-opportunity (ToO) request and were
scheduled on 2009 May 12 - 13 to coincide with the multi-wavelength
campaign.

\subsection{\it XMM-Newton EPIC-pn}

The EPIC-pn camera (Turner et al. 2001) was operated in burst mode, with the thin filter
because the high flux of the source ($>$ $1.3\times 10^{-8}$ erg s$^{-1}$ cm$^{-2}$) would result in strong pile-up.
In burst mode, it is arranged that the photon collection efficiency (the live time) is reduced to 3\% 
allowing observations of bright sources. After correcting for the live time, Cygnus\th X-2 produced 
a count rate of $\sim$3\th 000 count s$^{-1}$, very much less than the pile-up limit of 60\th 000 count 
s$^{-1}$. In burst mode, spatial information is only available in the x-direction, perpendicular to the 
CCD readout direction. This mode provides high time resolution of 7 $\mu$s and a moderate energy 
resolution of \hbox{E/$\Delta$ E} $\sim$ 40 at 7 keV. 

Analysis of the EPIC-pn data was carried out using the 
SAS 9.0.0 software and calibration data from April, 1 2010. 
A barycentric correction was applied to the event file using {\sc barycen} and also a correction for 
the charge transfer inefficiency (CTI) effect seen in EPIC-pn fast mode data using {\sc epfast}; without this
correction an unreal feature will remain in the spectrum at $\sim$2 keV. 

Lightcurves were extracted from the events file by selections made in the one-dimensional image 
using {\sc evselect}. During extraction, standard screening was applied to use single and double events
and remove known hot pixels. In  the pseudo one-dimensional image the y-axis for RAWY $<$ 180 represents 
time and a lightcurve can be derived from this. The part of the y-axis 
used was further restricted to RAWY $<$ 140 based on the tests made by Kirsch et al. (2006) in observations of 
the Crab pulsar and the Crab nebula, also using EPIC-pn data in burst mode. Using the Crab nebula as a calibration 
source, they showed that data between RAWY = 140 - 180 should not be used to prevent pulse pile-up affecting not 
only the slope, but also the normalization of the spectrum. Source data were extracted from a strip 25 pixels wide 
centred on the source at pixel 38. Background was obtained in two strips, each 5 pixels wide, one centred on 
pixel 5 and the other at pixel 62, as far from the source as possible on each side of the source.

To check for the presence of flaring particle background due to protons in the Earth's magnetosphere,
a lightcurve was extracted using all RAWX values in the 10 - 12 keV energy range with single events only 
(PATTERN = 0). Periods of increased background activity were detected in this way, but at the level of 0.3 count s$^{-1}$
and so totally negligible in comparison with the source intensity.

Various corrections were made using {\sc epiclccorr} including those for vignetting, bad pixels and the point
spread function; this program also corrected for the 3\% live time by scaling the intensity to the value
it would have for 100\% live time, and background was then subtracted.
A background-subtracted and deadtime-corrected lightcurve was extracted in the total energy band 0.2 - 10.0 keV
and also in several sub-bands 0.5 - 2.5 keV, 2.5 - 4.5 keV, 4.5 - 8.5 keV used for derivation of
hardness ratios as described below. 
 
Spectra were extracted with {\sc evselect} using single and double events. Response functions consisting 
of the response matrix file (rmf) and ancillary response file (arf) were built using {\sc rmfgen} and {\sc arfgen}.

\subsection{XMM MOS}

The MOS detectors could not be used because of the source brightness, and so both detectors MOS1 and
MOS2 were blanked across the central CCD so that no useful data on Cygnus\th X-2 were obtained.

\subsection{XMM RGS}

The standard mode of the Reflecting Grating Spectrometer (RGS; den Herder et al. 2001)
reads out eight CCDs of RGS\th 1 (via two nodes) and eight CCDs of RGS\th 2 (via a single node),
resulting in accumulation times of 4.8 s and 9.6 s, respectively. 
%Unlike for {\it Chandra}'s 
%Advanced CCD Imaging Spectrometer (ACIS) with an integration time of 2.85 ms in Continuous 
%Clocking (CC) mode (below), 
Photon pile-up in the dispersed spectrum can be an issue and we
therefore aimed with the RGS observation at the spectral range above 20 \AA,
which is not as well covered by the {\it Chandra} High Energy Transmission Grating Spectrometer (HETGS). 
The readout sequence of four CCDs only
on each RGS was optimised such that the frame time of the CCD receiving the highest flux on each RGS
was reduced to a quarter of the default value. This allowed us to monitor the oxygen edge at 
low count rates per frame time free of pile-up.

The RGS data were analysed with {\it XMM} SAS 9.0.0 software, following standard procedures:
{\sc rgsproc} was used to extract spectra and create response files, and {\sc rgslccorr} to 
create exposure-corrected light curves in the full 0.33 - 0.60 keV (20.7 - 37.6 \AA) band,
as well as in the 0.33 - 0.52 keV (23.8 - 37.6 \AA) band below the O edge, and in the 
0.55 - 0.60 keV (20.7 - 22.5 \AA) band above the the edge and associated O\,\textsc{i} 
resonance absorption lines.

\subsection{\it XMM OM}

The Optical Monitor (Mason et al. 2001) was used in ``Image Fast'' mode and the ultraviolet filter UVW1 was
used, this filter giving the best sensitivity in the desired wavelength range of 2500 - 4000 \AA. The B
filter was not used because of two bright stars in the field of view. The observation was divided into
exposures 4400 seconds long, giving coverage of almost all the X-ray observations and providing a
lightcurve with 10 s binning.

Data from the Optical Monitor (OM) were analysed using standard procedures. Firstly, tracking
corrections were applied using the tools {\sc omprep, omdrifthist} and {\sc omthconv}.
Flat-fielding was carried out using {\sc omfastflat}, and in the final step lightcurves were
produced using {\sc omregion}, {\sc evselect} for source and background selection, and
{\sc omlcbuild}.

\subsection{\it Chandra HETGS}

The {\it Chandra} observations with the High Energy Transmission
Grating spectrometer spanned about 67 ksec beginning at 13:00 UT on May 12 and had very good
overlap with the {\it XMM} exposures. Because of the extreme brightness of the source in combination
with the very narrow point spread function, strong photon pile-up is expected for normal
CCD frametime operations. Thus we chose the fastest available detector readout time (3.82 msec) provided
in continuous clocking (CC) mode as done in a previous {\it Chandra} observation of
the source (Schulz et al. 2009). 

All data were analysed using CIAO4.2 and the most recent CALDB products.
The first order count rates were quite similar during the two observations,
averaging about 220 count s$^{-1}$ leading to an average value for the counts/frametime/node
of $\sim 0.025$, and for the counts/frametime/pixel, a value well below 0.01. We are therefore not
worried about pile-up in the HETG spectra. Bright sources observed in CC mode have 
different charge transfer inefficiency properties and have to be treated
somewhat differently with respect to the normal timed event mode (see Schulz et al. 2009). 

\subsection{Radio observations}
Cygnus X-2 was observed using the e-EVN from 2009-May-12 23:52:22 UT
until 2009-May-13 12:54:03 UT at 5~GHz. The telescopes participating
were Jodrell Bank MkII, Knockin, Cambridge, Effelsberg, Torun, Yebes,
\begin{figure*}[!ht]                                                                         % Fig. 1
\begin{center}
\includegraphics[width=160mm,height=160mm,angle=270]{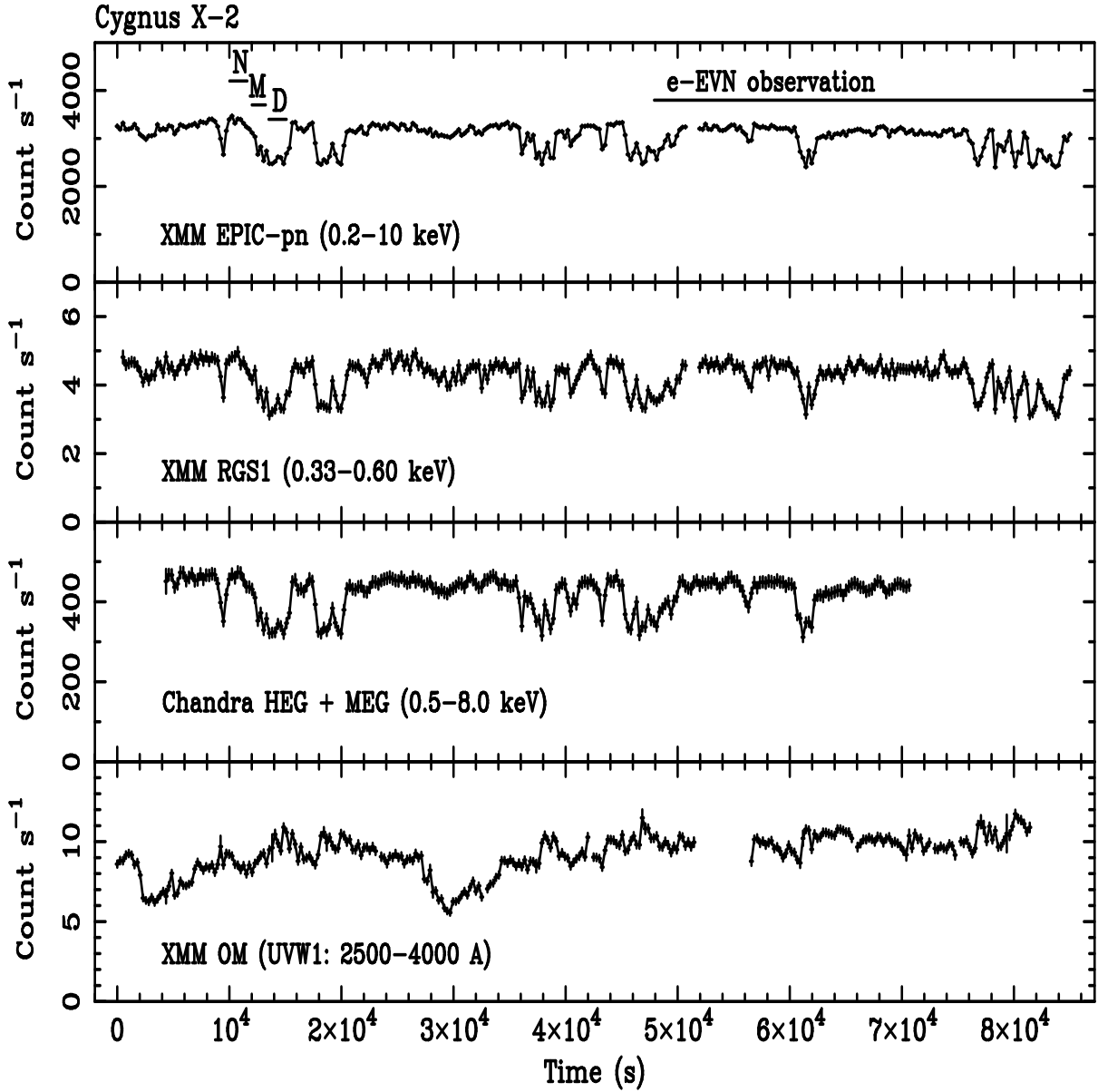}                              % was fig1_lc_4_256s
\caption{Upper panel: background-subtracted and deadtime-corrected EPIC-pn light curve of    % corrected version new_lc
Cygnus\th X-2 in the band 0.2 - 10.0 keV; spectra of non-dip, medium and deep dipping
were selected from the sections labelled ``N'', ``M'' and ``D''. Upper middle panel:
the corresponding 0.33 - 0.60 keV RGS1 lightcurve; lower middle panel: light curve of the {\it Chandra}
HETGS HEG + MEG instruments in first order; lower panel: the {\it XMM} Optical Monitor
lightcurve in the UVW1 band; all four lightcurves have 256 s binning.}
\label{}
\end{center}
\end{figure*}
Medicina, Onsala 25-m and Sheshan. Data were transferred in real time
from each antenna to the correlator using high-speed dedicated e-VLBI
network links\footnote{e-VLBI developments in Europe are supported by
the EC DG-INFSO funded Communication Network Developments project
'EXPReS', Contract No. 02662.}; connection rates of up-to 1024 Mbps were
sustained per antenna, giving a maximum bandwidth of 256~MHz. The field
of view was centred on $\alpha$ = $21^h 44^m 41^s.2, \,\delta =
38\degmark 19^{'} 18^{``}.0$ and phased reference to the calibrator
source J\th 2134+4050. Data were then reduced using the standard
\textsc{aips} VLBI algorithms and pipelined using EVN scripts written in
\textsc{parseltongue} (i.e. a python wrapper for classical \textsc{aips}).

\section{Radio results}

The target was not detected. The noise level of the image was $\sim$30 $\mu$Jy, hence the source 
had a five sigma upper limit of $\sim$150 $\mu$Jy per beam; the source was therefore in a
radio-quiet state. Hjellming et al. (1989) presented VLA results as part of the 1988 multi-wavelength
campaign on Cygnus\th X-2. Detailed correlations of radio flux at 1.49, 4.9 and 8.4 GHz with 
X-ray intensity and X-ray hardness showing that the source had a radio flux of 2 - 5 mJy when 
the source was on the horizontal or upper normal branch, but fell to a low level $<$ 1 mJy on 
the lower normal and flaring branches, which was designated 
\begin{figure*}[!ht]                                                                   % Fig. 2
\begin{center}
\includegraphics[width=160mm,height=160mm,angle=270]{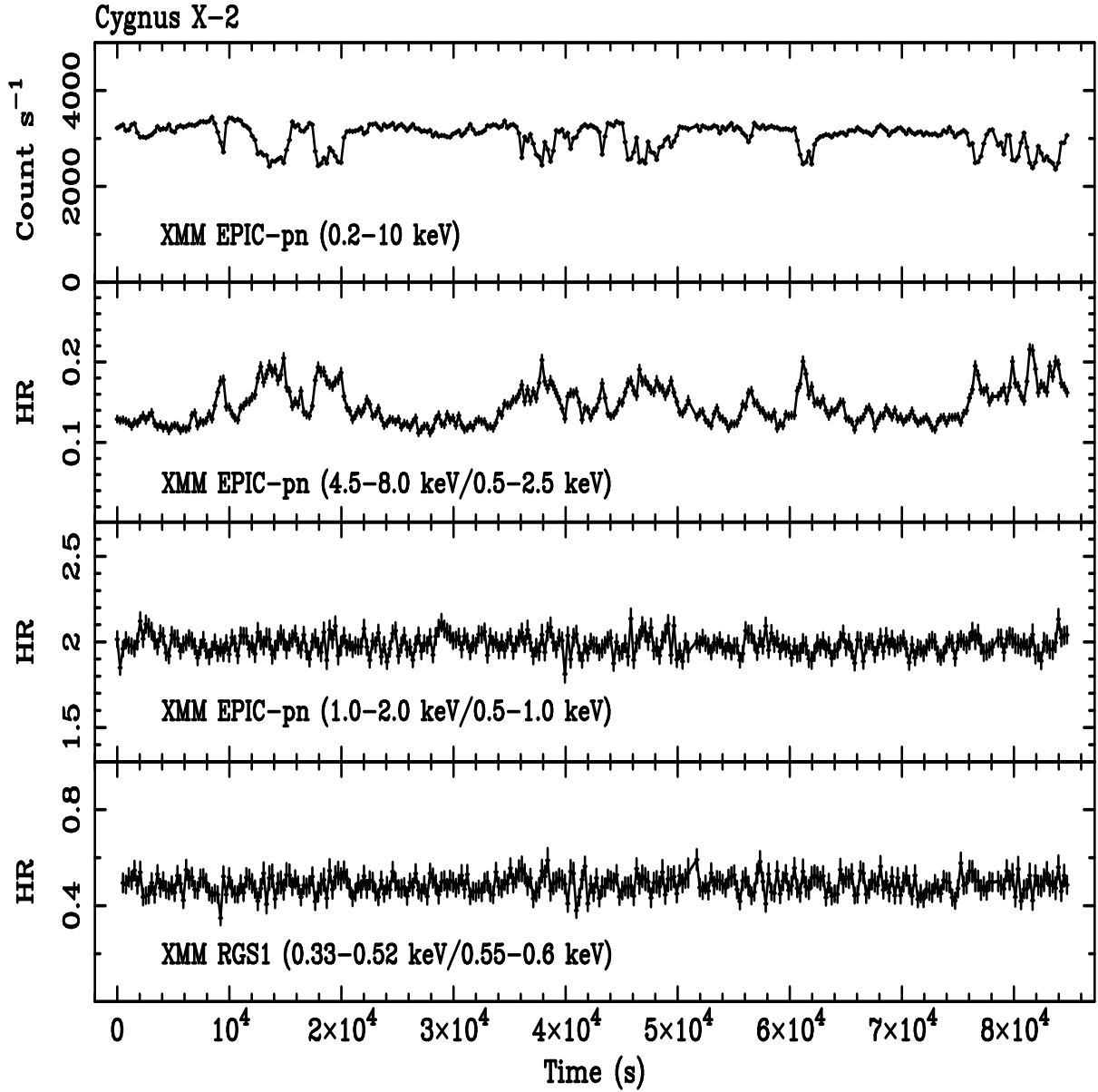}                        % was fig3_HR_4_256s
\caption{Hardness ratios defined in several energy bands: upper middle panel:
using wide EPIC-pn bands; lower middle band: using EPIC-pn bands below 2 keV;
lower panel: using the RGS1 band. The upper panel gives the 0.2 - 10 keV EPIC-pn lightcurve
for comparison.}
\label{}
\end{center}
\end{figure*}
as a radio-quiet state. The present 
X-ray results indicate the source to be in a stationary position on the Z-track at or close to the 
soft apex, and so the lack of radio detection is consistent with this. The radio results have 
been published in Astronomer's Telegram 2052 by Rushton et al. (2009)

\section{X-ray results}

In Fig. 1 (upper panel) we show the {\it XMM-Newton} EPIC-pn lightcurve with 256 s binning
together with the corresponding RGS1 lightcurve (upper middle panel), the {\it Chandra} HETG
lightcurve (lower middle panel) and the OM lightcurve (lower panel).
It can be seen that the {\it XMM} observation spanned 86 ksec, while
the {\it Chandra} observation was somewhat shorter at 67 ksec. All X-ray lightcurves clearly display reductions in intensity 
of up to 25\%, having the appearance of absorption dips as seen in the dipping class of LMXBs. 
The dipping is structured but seems to have a consistent lower level, 
or maximum depth of dipping. In addition to this variability, there is a continuous small decrease in 
the non-dip intensity across the observation from $\sim$3400 to $\sim$3300 count s$^{-1}$. 

%not an X-ray result...
%Dipping may also be seen in the Optical Monitor lightcurve which is the first 
%detection of dipping in the optical/UV. In addition there are other stonger decreases in OM 
%intensity of apparently different nature, at times of 0.3 ksec and 30 ksec and the OM results 
%will be discussed in Sect. 7.5 after presentation of the X-ray results.

\subsection{Hardness ratios}

The source changes were investigated by the use of several X-ray hardness ratios from the EPIC-pn 
and RGS instruments
as shown in Fig. 2. The upper panel shows the EPIC-pn 0.2 - 10 keV lightcurve for comparison 
with the upper middle panel giving a hardness ratio  defined as the ratio of counts per second
in the band 4.5 - 8.0 keV to that in the band 0.5 - 2.5 keV, these bands being chosen to allow
comparison with {\it Chandra} data, which do not extend to 10 keV. The lower middle panel shows 
a  hardness ratio from EPIC-pn defined using the bands 1.0 - 2.0 keV and 0.5 - 1.0 keV, in the low 
energy part of the spectrum and the lower panel shows a similar ratio from RGS1 using the
bands 0.33 - 0.52 keV and 0.55 - 0.60 keV. The hardness ratio based on the wide PN band 
shows a factor of two hardening of the spectrum during the dips which 
is the behaviour demonstrated in many dipping LMXB sources, 
reflecting the tendency for absorption to remove the lower energies; however, this is not always 
the case as the spectrum consists of two emission components: the neutron star and the ADC, 
which may have different degrees of absorption (Church et al. 1997). The hardness ratios
defined in bands below 2 keV remarkably show no change in hardness, which is a result of
the nature of the absorbing process, which will be discussed in Sects. 4.3 and 5.2.

The source variations were investigated further by displaying the {\it XMM} EPIC-pn data 
on a hardness-intensity diagram as shown in Fig. 3. The same hardness ratio was used as in Fig. 2
(upper middle panel). Sections of data in the lightcurve were identified in the hardness-intensity 
diagram by making a number of selections of short sections of data.

\begin{figure}[!ht]                                                        % Fig. 3
\begin{center}
\includegraphics[width=80mm,height=80mm,angle=270]{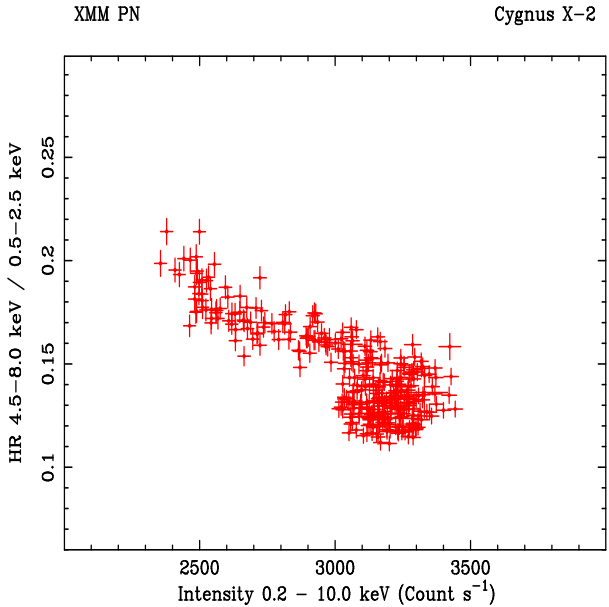}              % was ztrack_lccorr_3_256s 
\caption{Hardness-intensity diagram for the {\it XMM} PN data.}
\label{}
\end{center}
\end{figure}

This revealed that the non-dip data form an approximately circular soft region
at the right hand side of the diagram. Data of decreased intensity form
the trail extending to lower intensities and harder parts in Fig. 3. It can be seen that
the source did not move on its Z-track, but remained in a fixed position during the observations.
Spectral fitting described below shows that the decreases of intensity are indeed absorption
events, i.e. X-ray dips with column density increases, as opposed to the possibility
of this trail being part of the horizontal branch of the Z-track.

\subsection{Spectral fitting of Epic-pn spectra}

Spectra were extracted by selection on times and intensities from the EPIC-pn lightcurve 
corresponding to non-dip, medium dip and deep dip data (labelled N, M and D in Fig. 1).
Using the section of strong dipping near the start of the observations (at 10 - 15 ksec), 
a non-dip spectrum was selected having intensity $>$ 3400 count s$^{-1}$, i.e that part of the 
EPIC-pn lightcurve at times between 10 - 12 ksec. An intermediate dip spectrum was taken with
intensities between  2970 - 3120 count s$^{-1}$ corresponding to times of $\sim$12 - 13 ksec, and 
a deep dip spectrum at intensities $<$ 2460 count s$^{-1}$, at times of $\sim$13 - 16 ksec. The 
non-dip spectrum was obtained as close to the dip data as possible to avoid any drift in source intensity.
Spectra were re-grouped to effectively oversample the spectrum by a factor of three at each energy.
This required primitive channels to be binned in groups of five below 1.0 keV, in groups of eight
between 3 - 4 keV, and in tens between 7 - 8 keV etc. Spectra were fitted in the range 0.6 - 12.0 keV.

These spectra were fitted using the extended ADC emission model so that the non-dip spectrum was 
fitted by a model of the form {\sc abs}$\ast${\sc (bb + cut)}, where {\sc abs} is the Galactic absorption,
{\sc bb} the neutron star blackbody emission and {\sc cut} the Comptonized emission of an extended ADC 
(Church \& Ba\l uci\'nska-Church 2004). In fitting, the {\it Xspec} model {\sc phabs} was used as a 
more accurate form than the generally used {\sc wabs} model. A background spectrum was 
obtained from the strips at each side of the CCD (Sect. 2.1) using the whole observation, and allowance 
made for the difference in area using {\sc backscal}. In the Z-track sources, the Comptonization high
energy cut-off is relatively low at about 5 keV, which does not allow determination of the
power law photon index given the narrow available energy range. Accordingly, the index was fixed
at the physically reasonable value of 1.7 (Shapiro et al. 1976) as we have previously carried out as standard 
(Church et al. 2006). 

In preliminary testing, it became evident that dipping consisted of absorption of the Comptonized
emission only, and to demonstrate this, the non-dip and deep dip spectra were fitted simultaneously by
a model in which each emission term had a scaling factor: {\sc k1}$\ast${\sc bb} + {\sc k2}$\ast${\sc cut}. 
For the non-dip spectrum, k1 and k2 were frozen at unity. Fitting showed that during dipping, k1 remained 
closely equal to unity: k1 = $1.01\pm 0.05$  at 90\% confidence but k2 decreased to $\sim$0.60, 
indicating that the Comptonized emission was removed while the blackbody was not affected. This contrasts
markedly with spectral evolution in the dipping LMXB in which the blackbody is absorbed with a much 
higher column density that the Comptonized emission, indicating that the neutron star emission
is subject to absorption along a path of denser material through central parts of the bulge in
the outer disk (Church et al. 1997). Absorption could be seen to take place at energies below
5 keV, demonstrating that photoelectric absorption is taking place. The neutron star blackbody
spectrum, however, peaks at $\sim$3 keV so that its flux at 1 keV is relatively small comprising only 10\%
of the total flux. At 0.5 keV this falls to 1\%, and it is clear that absorption of the blackbody
cannot explain the dipping and that the predominant absorption process is absorption of the
Comptonized emission (see Fig. 4).

From this point onwards, fitting was carried out using a model of the form 
{\sc abs}$\ast${\sc (bb + pcf}$\ast${\sc cut)} where {\sc abs} is Galactic interstellar absorption and
{\sc pcf} is a progressive covering factor as appropriate to removal of extended emission by an extended 
absorber passing across the line-of-sight. In analysis of the dipping LMXB, spectral evolution in dipping 
could be well-described by a model in which the neutron star blackbody was absorbed and the ADC emission 
was progressively covered by absorber (e.g. Church et al. 1997), i.e. a model 
{\sc abs}$\ast${\sc (abs}$\ast${\sc bb + pcf}$\ast${\sc cut)} representing the effects
of the absorbing bulge on the outer disk passing across the neutron star, and also progressively
overlapping the extended ADC. The difference in the present case is the lack of strong absorption of 
the neutron star blackbody emission.

The non-dip, intermediate dip and deep dip spectra were fitted with this model. It was possible
to fit the non-dip spectrum first and then apply the emission parameters found to the dip spectra,
or alternatively, to fit the three spectra simultaneously. Both techniques provided good fits, and
we present here results of simultaneous fitting. In this case, the progressive covering factor was
frozen at zero for the non-dip spectrum, but was free for each of the dip spectra. Emission parameters
were chained between all three spectra, forcing them to be equal. With the above continuum model, 
a broad excess could be seen in the residuals at 1 keV and a weak feature at $\sim$6.6 keV. Testing 
showed that the 1 keV feature was reduced in flux in dipping and could be modelled in dipping by 
including the Gaussian line inside the partial covering factor, this suggesting that the emission 
originated in the ADC. The Fe feature was weaker and its behaviour in dipping could not be constrained.

A good simultaneous fit was obtained having $\chi^2$/d.o.f. = 633/588 = 1.08 and when each spectrum 
was individually examined, the fit was good in each case (see Table 1).
%($\chi^2$/d.o.f. = 279/244. 291/245 and 
%241/245 for non-dip, intermediate and deep dipping).

The blackbody emission was found to have a temperature $kT_{\rm BB}$ of 1.29$\pm$0.07 keV,
and a blackbody radius $R_{\rm BB}$ of $(9.6 \pm 1.1)\times (d/8\, kpc)$ km for a source distance of 8 kpc.
The Comptonized emission was well described by a cut-off power law with power law index 1.7 and
cut-off energy 8.9$\pm$1.9 keV; all uncertainties are quoted at 90\% confidence. This cut-off energy 
is higher than the values of 4 - 5 keV that we have previously obtained for the Z-track sources, presumably 
because we can only fit up to 12 keV in the present case. The Galactic column density $N_{\rm H}^{\rm Gal}$ was found to be 
$(3.6\pm 0.1)\times 10^{21}$ atom cm$^{-2}$ assuming the cross sections of Verner et al. (1996) and
the abundances of Wilms et al. (2000), substantially higher
than the value of $(2.3\pm 0.5)\times 10^{21}$ atom cm$^{-2}$ of Juett et al. (2004).
The total 1 - 10 keV luminosity of the source was found to be $L_{\rm Tot}$ = $9.83\times 10^{37}$ erg s$^{-1}$
for a distance of 8 kpc (Cowley et al. 1979).

\begin{figure}[!ht]                                                            % Fig. 4
\begin{center}
\includegraphics[width=80mm,height=80mm,angle=270]{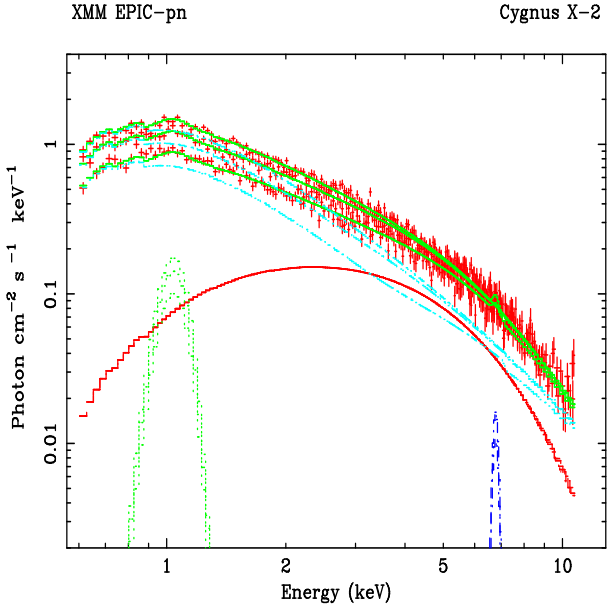}                  % was m11b;  improved version fig4_spec_col
\caption{Best fit to the non-dip, intermediate dip and deep dip spectra showing in each case the data, 
total model, blackbody and Comptonized emission components and the emission lines at 1 and 6.6 keV.
It can be seen that in the best-fit model, the blackbody curve peaking at 3 keV does not change
during dipping as it is not subject to absorption.}
\label{}
\end{center}
\end{figure}

To better obtain the line parameters, a spectrum was constructed containing all EPIC-pn non-dip data 
(excluding the features at $\sim$3 and $\sim$30 ksec (Sect. 4.5), and using this the 1 keV feature 
was approximately described by a Gaussian line at 1.03$\pm$0.03 keV and width $\sigma$
= 0.09$\pm$0.03 keV with equivalent width of 28 eV. Close examination of the Fe feature
revealed structure which could be fitted by two Gaussian lines; however, it was necessary to freeze 
each $\sigma$ at 30 eV to represent the width seen in the residuals. The line energies were
found to be 6.61$\pm$0.04 keV and 6.88$\pm$0.05 keV with EW of 18 and 17 eV indicative of states Fe~XXV 
and Fe~XXVI in agreement with the values obtained by Schulz et al. (2009) of 6.662$\pm$0.018 and 6.919$\pm$0.035 keV.
Schulz et al. resolved the Fe XXV He-like triplet and the Fe XXVI H-like resonance line with moderate 
line widths that correspond to 3450 km s$^{-1}$ and 1120 km s$^{-1}$, respectively. This indicated 
formation at a radius much greater than 10$^9$ cm from the neutron star.
% Schulz et al. identified three components of Fe XXV as resonance, intercombination and forbidden lines
% while Fe XXVI was the resonance line. The double-peaked Fe XXV feature with Doppler velocities of
% 3450 km $s^{-1}$ indicating formation at greater than $1.8\times 10^9$ cm from the neutron star.
Done et al. (2002) claimed that the broad Fe feature in {\it Ginga} observations of Cyg\th X-2 was well 
described as associated with reflection in the accretion disk of a central Comptonizing region around 
the neutron star. Similarly, Shaposhnikov et al. (2009) fitted {\it Suzaku} data with a disk reflection model.
The HETG spectra of Cyg\th X-2 so far only show contributions from ionized Fe lines of moderate
broadening. Contributions from neutral Fe fluorescence can be ruled out. Fe fluorescence is only seen 
in a minority of LMXB; the HMXB Cyg\th X-3 in {\it Chandra} clearly exhibits \hbox{Fe K-$\alpha$} fluorescence 
{\it and} the hot photoionization lines Fe XXV and Fe XXVI (Torrej\'on et al. 2010). The Chandra spectra of Cyg X-2
indicate emissions from the outer part of the system rather than reflection from the inner disk.
%However, our results resolving the feature into two lines and the work of Schulz et al. (2009) resolving
%Fe XXV into a He-like triplet rule out Fe fluorescence and indicate line origin in an extended ADC.
%In our case, the complete absence of the fluorescence line shows that
%reflection does not take place.} 

The intermediate dip and deep dip spectra were well described by the above emission parameters
together with the partial covering factors and column densities shown in Table 1. The covering factor
term {\sc pcf} may be written as $f\ast${\sc abs} + (1 - $f$), i.e. a fraction $f$ of the
Comptonized emission is subject to additional absorption with column density $N_{\rm H}$ (as
shown in Table 1) while the remainder of the emission $(1 - f)$ is not overlapped by absorber.
It can be seen that the covering factor rises to 40\% in deep dipping with a high column density
of $45\times 10^{22}$ atom cm$^{-2}$ which strongly removes ADC emission below 5 keV. This can be
seen in Fig. 4, which shows the unfolded spectra and the best-fit models obtained by simultaneous 
fitting. 

\begin{table}
\begin{center}
\caption{Spectral evolution in dipping. Column densities are given in units of $10^{22}$ atom cm$^{-2}$.}
\begin{minipage}{80mm}
\begin{tabular}{lrrr}
\hline \hline\\
$\;\;$ Spectrum  & $f$                  & $N_{\rm H}$            & $\chi ^2/d.o.f.$ \\
\noalign{\smallskip\hrule\smallskip}
Non-dip          & 0.0                &0.0                     & 223/194 \\
\noalign{\smallskip}
Intermediate dip &0.182$\pm$0.019     &13$^{+13}_{-6}$     & 187/192 \\
\noalign{\smallskip}
Deep dip         &0.421$\pm$0.015     &45$^{+20}_{-12}$         & 223/192 \\

\noalign{\smallskip}\hline
\end{tabular}\\
% text
\end{minipage}
\end{center}
\end{table}

Finally a check was made of whether the neutron star blackbody emission could be subject to
a small amount of photoelectric absorption in addition to the strong absorption of the
Comptonized emission. This was done by adding an additional column density to the blackbody:
{\sc abs}$\ast${\sc bb}, frozen at zero for the non-dip spectrum, but free in the two dip spectra.
Free fitting showed that this extra column density remained at zero in both intermediate and
deep dip spectra (in deep dipping $N_{\rm H}$ = $0^{+1.7}_{-0.0}\times 10^{22}$ atom cm$^{-2}$).
There was no change in the quality of fit with $\chi^2$/d.o.f. = 633/586. 
This may be contrasted with fitting the dipping group of LMXB in which the Comptonized emission
is progressively covered, often reaching 100\% covering, but in which the neutron star blackbody
is also absorbed, and it is typically found that $N_{\rm H}^{\rm BB}$ is very high
having a {\it lower limit} of the order of $600\times 10^{22}$ atom cm$^{-2}$ (e.g. Church 
et al. 1997). It is clear that in the present case, absorption of this component is negligible.

%\begin{figure*}[!ht]                                                                   
%\begin{center}                                                             % Fig. 4
%\includegraphics[width=160mm,height=160mm,angle=270]{page1}                % was cygx-2_pdhetg_10881_flux  page1
%\caption{}
%\label{}
%\end{center}
%\end{figure*}

\subsection{XMM RGS results}

The RGS exhibited a similar lightcurve to those of the EPIC-pn and the {\it Chandra} gratings,
clearly showing the X-ray dips. Spectral fitting of the available RGS band (0.33 - 0.60 keV)
to determine spectral evolution in dipping cannot be sensibly carried out because of the 
narrowness of the band.
%If a model of the form {\sc abs}$\ast${\sc bb + cut} found to describe the EPIC-pn spectrum well were used, it
%would not be possible to determine the Comptonization high energy cut-off, or the power law index
%or the normalization because these require data in the band 1 - 10 keV. The blackbody contribution
%in the RGS band is almost zero. Thus if the power law parameters cannot be found, neither can
%the Galactic column density. However, it can be demonstrated that the RGS data are compatible with
%the progressive covering of the Comptonized emission found by spectral fitting of the EPIC data.
However, it is possible to investigate the effect of dipping on the oxygen edge. In Fig. 2 
(lower panel) we show a hardness ratio from RGS data defined as the ratio of counts in 
the bands 0.33 - 0.52 keV/0.55 - 0.60 keV, i.e. bands on either side of the oxygen edge. Thus, these energy 
bands indicate not only any hardness change in the RGS band, but also whether there is any increased
absorption in the oxygen edge at the time of dipping. No change at all is evident, which is
consistent with the EPIC spectral fitting results. This fitting showed that the X-ray 
emission in dipping consists of the covered part of the emission, which is subject to a high column 
density of absorber, plus the uncovered emission, which is not absorbed at all. In the RGS band 
the flux of the covered emission is negligible and the emission detected consists entirely of the
uncovered emission, which has essentially the non-dip spectrum scaled down by a factor (1 - $f$).
This result is confirmed by the hardness ratio in Fig. 2 obtained from EPIC data using counts
in the bands 1.0 - 2.0 keV/0.5 -1.0 keV, which also shows no change during dipping, as expected. In fact,
the EPIC-pn spectrum in non-dip divided by the dip spectrum is constant below 2.0 keV demonstrating
that the total removal of the covered emission produces an energy-independent reduction in the
spectral flux below 2 keV, and thus the low energy hardness ratios in Fig. 2 are accurately constant in dips.

Spectral fitting was carried out to determine absorption in the oxygen edge.
Because of the reduced count statistics compared with the EPIC-pn, it was necessary
to use longer sections of data and a dip spectrum was accumulated consisting of all degrees of
dipping (25 ksec exposure) and a spectrum consisting of all non-dip data (58 ksec). A 
parameterization of the local continuum was used consisting of two power laws. The first order
spectra of RGS1 and RGS2 were rebinned to S/N = 15 in the band 20.8 - 37.3 \AA. The absorption
edge of neutral oxygen and the fine structure due to the related series of resonance
absorption lines, in particular the strong 1s--2p transition of neutral oxygen at $\sim$23.5 \AA,
were modelled by a local absorption model ({\sc tbnew}) assuming the cross  sections of Verner 
et al. (1996) and the abundances of Wilms et al. (2000). The non-dip spectrum in the neighbourhood
of the edge between 20 - 28 \AA $\,$ together with the model is shown in Fig. 5. Fitting in this range with
a single power law for the continuum, 
we find an optical depth in the edge equivalent to a column density $N_{\rm O}$ = $(1.50\pm0.04)\times10^{18}$ 
O-atom cm$^{-2}$ and $(1.52\pm0.04)\times10^{18}$ cm$^{-2}$ for the non-dip and integrated dip spectra.
However, the systematic error in the column density derived from a local fit to the continuum may exceed
the statistical uncertainty. Thus, modelling the 20 -- 38 \AA $\,$range, needing two powerlaws for the continuum,
we obtain $N_{\rm O}$ = $1.75^{+0.07}_{-0.08}\times 10^{18}$ O-atom cm$^{-2}$ and $1.76^{+0.07}_{-0.08}\times10^{18}$ 
cm$^{-2}$ for the non-dip and dip spectra.
\begin{figure}[!ht]                                                                      
\begin{center}                                                                      % Fig. 5
\includegraphics[width=80mm,height=80mm,angle=0]{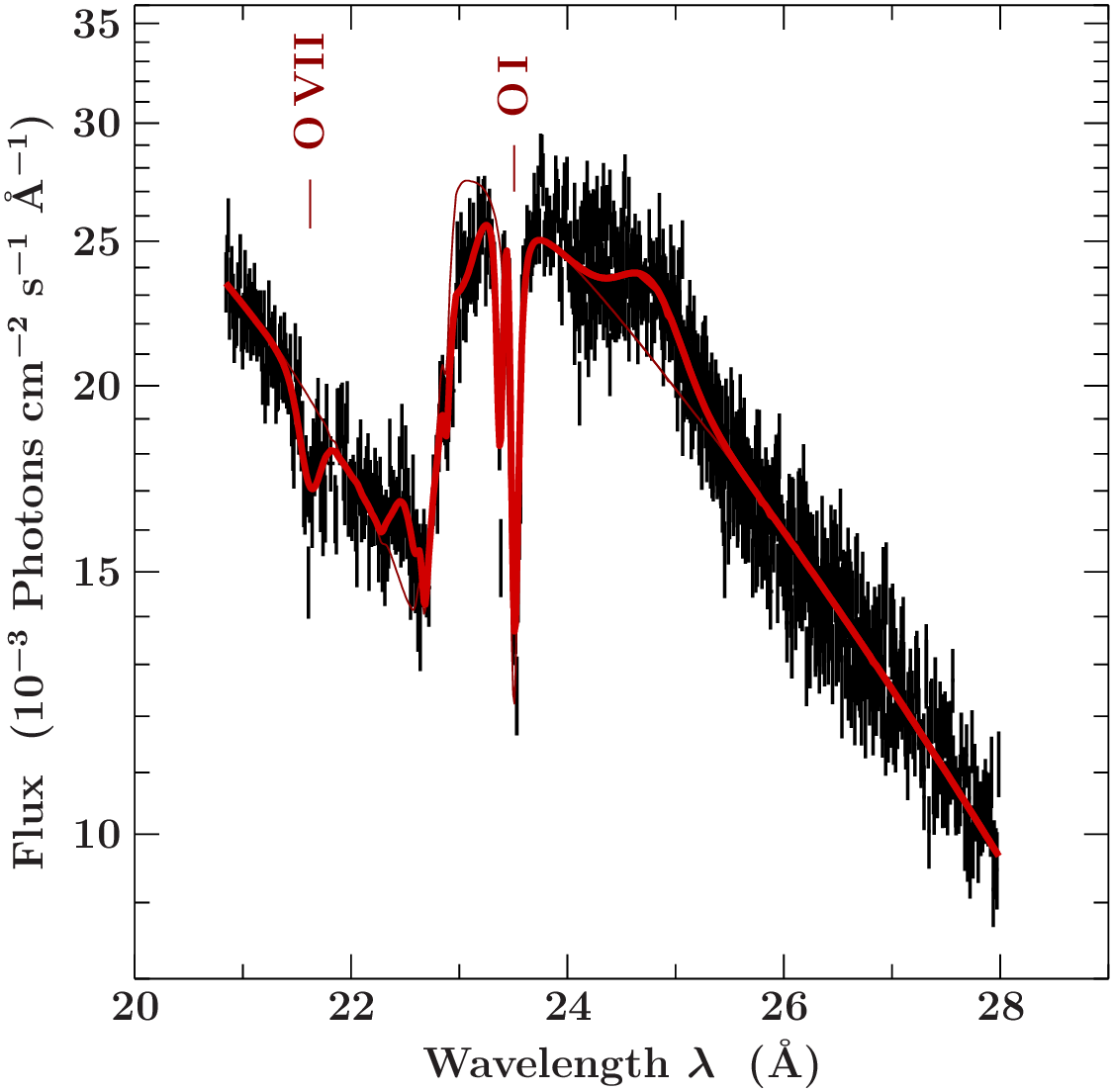}                         % was RGS_non-dip_20-28A
\caption{First order RGS spectrum in the region of the oxygen edge, together with the best-fit model.}
\label{}
\end{center}
\end{figure}
Assuming an abundance of oxygen in the interstellar medium of $4.9{\times}10^{-4}$
(Wilms et al. 2000), this latter oxygen
column corresponds to an equivalent hydrogen column
${N_H}$ of $3.57^{+0.14}_{-0.17}\times 10^{21}$ atom cm$^{-2}$ in non-dip, the value remaining unchanged 
in the integrated dip spectrum ($3.58^{+0.14}_{-0.17}\times 10^{21}$ atom cm$^{-2}$). Thus, this fitting 
supports the EPIC result that
in dipping only uncovered emission is seen and $N_{\rm H}$ reflects interstellar absorption.
The above values of $N_{\rm H}$ are in excellent agreement with that found in the EPIC-pn.

%COMMENT:
%Unless we are not using these detection to add to our
%dip/non-dip paradigm, I  think they are a distraction.
%I am sure we will find a future forum for these
%discoveries.!!
%We also detect additional absorption from O\,\textsc{ii} and O\,\textsc{iii}
%at 23.37\,\AA{} and 23.0\,\AA{} (see also Juett et al., 2004)
%as well as the recombination line of O\th\textsc{vi} at 21.61 \AA.
%In addition, narrow absorption lines at 30.99 \AA $\,$  and 31.3 \AA $\,$ were seen,
%probably Li-like Si XII and Ca XI, and in addition to these, weak broad emission features
%at 22,5, 24.8 and 29.1 \AA.  {\bf Mention these ?}

\subsection{{\it Chandra} grating results}

We now present results from the {\it Chandra} gratings for the present observation,
and also compare these in Sect. 4.4.1 with all previous {\it Chandra} grating results for Cygnus\th X-2.
We can use the HETGS data to investigate photoelectric absorption during persistent
emission and during dips. We detect absorption edges for Mg K, Ne K, Fe L, and 
O K at 9.48\AA $\,$(1308 eV), 14.29\AA $\,$(867 eV), 17.51\AA $\,$(708 eV), and 
22.89\AA $\,$(542 eV) (Juett et al. 2004, 2006). 
Since the \emph{Chandra} data were taken in CC graded mode, we cannot model
the spectral continuum as we did for the pn-spectra, but need to defer
to a local continuum modelling. The lightcurve from the summed \emph{Chandra} first
order grating data is shown in the lower middle panel of Fig. 1. It excellently matches the 
pn-lightcurve from 4 to 71 ksec. Note, that on top of heavy dipping activity,
the overall persistent flux decreases by about 5$\%$ over the entire exposure and 
instead of a simple cut by rate, we hand-selected and accumulated about 13 ksec of
persistent events. Similarly, the dip minima vary considerably and again
we selected and accumulated about 10 ksec of dip events from the bottoms of the dips.    
This selection results in a further reduction of the
already reduced statistics with respect to the EPIC-pn,
and we had to accumulate and add as many persistent and
dipping intervals as possible throughout the lightcurve.
\begin{figure*}[!ht]
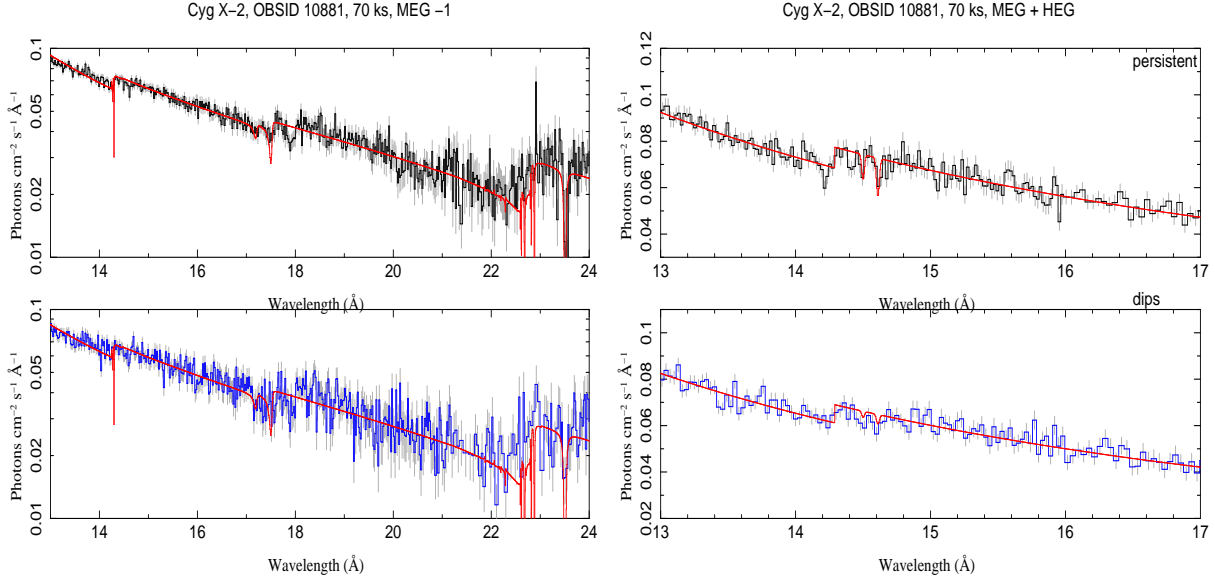
                                                                   % Fig. 6
\begin{center}
\includegraphics[width=80mm,height=80mm,angle=0]{15931fig6a}              
\includegraphics[width=80mm,height=80mm,angle=0]{15931fig6b}
\caption{Left: Broadband fit using a multi-component continuum model and the {\sc tbnew} high-resolution 
absorption function to the MEG -1st order spectra of the persistent (top) and dip (bottom) emissions.
The bandpass includes the Ne K edge at 14.29~\AA (867 eV), the Fe L II and III edges at 17.2~\AA (721 eV) 
and 17.5~\AA (708 eV), and the O K edge at 22.89~\AA (542 eV). Right: The local fit to the Ne K edge using 
a simple power law and an edge function with an energy fixed at 14.29~\AA.}
\label{hetg_edges}
\end{center}
\end{figure*}

The left side of Fig. 6 shows broadband fits to the non-dip
data (upper panel) and dip data (lower panel) using a multi-component 
continuum model consisting of a power law and a blackbody component and the 
{\sc tbnew} high-resolution absorption function. Here we only use
spectra from the MEG -1st order due to some hot pixel contamination in the 
O K edge in the MEG +1st order, which in CC-mode is difficult to remove.
The fit in the persistent emission provides fair matches to the edges with
a hydrogen equivalent of $(2.5\pm 0.3)\times10^{21}$ cm$^{-2}$ for O K and 
Fe L, the one at Ne K appears higher at (3.0$\pm0.3)\times10^{21}$ cm$^{-2}$.
The value from the oxygen edge is smaller than seen in the {\it XMM} RGS
and it is possible that this is due to residual cross-calibration uncertainties.
The fits to the persistent and dip emission are practically identical as seen in the RGS, 
and the dip spectra do not seem to require adjusted optical depths, and we emphasize 
the surprizing nature of these results given the clear reduction of continuum intensity 
in dipping. 

The optical depth at Ne K is much better determined because it provides the best statistics. 
Here we can use all the MEG data and the contributions from the HEG. We thus focus on local 
fits to the optical depth, and for this we simply apply an almost flat power law plus an edge 
function, which we fix at the expected rest value of the edge at 867 eV $\,$(14.29\AA). The 
results are shown in Fig. 6 (right panel).
\begin{figure}[!h]                                                                   % Fig. 7
\begin{center}
\includegraphics[width=76mm,height=76mm,angle=0]{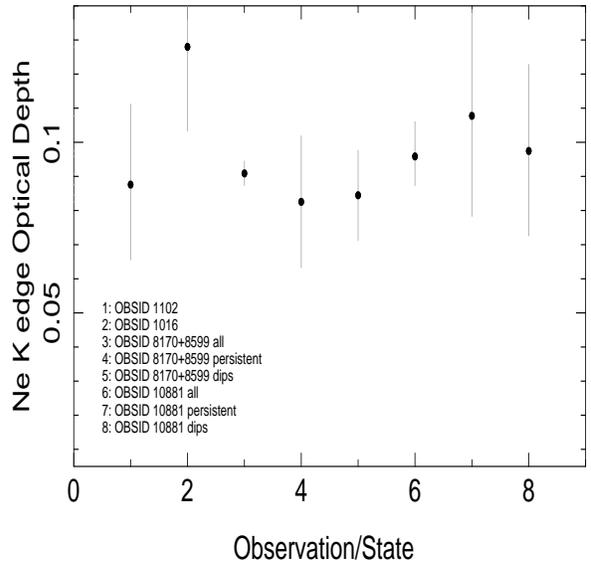}                          % was needge_depths.ps
\caption{\bf Optical depth measurements of Ne K edges for all HETGS                  % new version with extra point is
observations in the \emph{Chandra} archive.}                                         % needge_depths01.ps
\label{hetg_depths}
\end{center}
\end{figure}
The scatter in the persistent and the dip spectrum is quite comparable and it is already visible
that the edge depths are very similar. As described above, the equivalent hydrogen column density
derived from the optical depth of 
\tabcolsep 3.5 mm
\begin{table*}[!ht]
\begin{center}
\begin{minipage}{160mm}
\caption{Ne K optical depth and equivalent hydrogen column $N_H$ = $N_{\rm Ne}/8.71\times 10^{-5}$
from local fits during persistent emission and dipping.
{\it All} means all integrated data, {\it non-dip} depicts flat and dip-free parts
of the persistent emission, and {\it deep dip} the parts at the bottom of dips.}
%\begin{minipage}{76mm}
\begin{tabular}{lrrrr}
\hline \hline\\
state     & date (Obsid)        & $\tau_{\rm Ne}$       & $N_{\rm Ne}$    & $N_{\rm H}$\\
          &                     &                       & (10$^{17}$ cm$^{-2}$)  & (10$^{21}$ cm$^{-2}$)\\
\noalign{\smallskip\hrule\smallskip}
All       & 1999 Sep 23 (1102)  & $0.087\pm 0.023$      & $2.41\pm0.63$  & $2.77\pm0.72$\\
\noalign{\smallskip}
All       & 2001 Aug 12 (1016)  & $0.128\pm 0.025$      & $3.52\pm0.69$  & $4.04\pm0.80$\\
\noalign{\smallskip}
All       & 2007 Aug 23, 25     & $0.091\pm 0.004$      & $2.50\pm0.10$  & $2.87\pm0.11$\\
\noalign{\smallskip}
Non-Dip   & (8170+8599)         & $0.083\pm 0.019$      & $2.27\pm0.53$  & $2.61\pm0.61$\\
\noalign{\smallskip}
Deep dip  &                     & $0.084\pm 0.013$      & $2.32\pm0.36$  & $2.67\pm0.42$\\
\noalign{\smallskip}
All       & present obs         & $0.096\pm 0.009$      & $2.64\pm0.26$  & $3.03\pm0.30$\\
\noalign{\smallskip}
Non-Dip   & 2009 May 12         & $0.108\pm 0.030$      & $2.96\pm0.83$  & $3.40\pm0.96$\\
\noalign{\smallskip}
Deep dip  &  (10881)            & $0.097\pm 0.025$      & $2.68\pm0.69$  & $3.08\pm0.79$\\
\noalign{\smallskip}\hline
\end{tabular}\\
% text
\end{minipage}
\end{center}
%Abundance of Ne is          alog( 12.0 - 8.14) ratio =  7244  measured ratio = 1149
\end{table*}
the Ne K edge appears slightly higher than for other edges. In order to investigate this we 
analysed the Ne K edge region in several ways. In addition to the persistent and dip segments, 
we also determined the edge depth for the entire observation. 
This gave equivalent $N_{\rm H}$ values of $(3.40\pm 0.96)\times 10^{21}$ atom cm$^{-2}$ for the non-dip data,
$(3.08\pm 0.79)\times 10^{21}$ atom cm$^{-2}$ for deep dipping and
$(3.03\pm 0.30)\times 10^{21}$ atom cm$^{-2}$ when all data were added.
The non-dip value is in good agreement with the $N_{\rm H}$ values from EPIC-pn and the RGS.

\subsubsection{Comparison with previous {\it Chandra} observations of Cyg\th X-2}

We also repeated the
same procedure using previous HETGS data of Cyg X-2, i.e. obsid 1102 for 29 ksec (Juett et al. 2004),
obsid 1016 for 14.6 ksec, and obsids 8170 and 8599 for 134 ksec (Schulz et al. 2009). The results are
shown in Table 2 and Fig. 7, and the results for the present observation are added to these.
Our latest values agree with the previous values within uncertainties.

Juett et al. (2004) derived a Galactic column density for Cygnus\th X-2 from the oxygen edge
of $N_{\rm H}$ = $(2.3\pm 0.5)\times 10^{21}$ atom cm$^{-2}$, and Juett et al. (2006) obtained
a column from the Ne edge $N_{\rm Ne}$ of $(2.3^{+0.9}_{-0.3})\times 10^{17}$ atom cm$^{-2}$
equivalent to a hydrogen column density of $(2.64^{+1.0}_{-0.34})\times 10^{21}$ atom cm$^{-2}$
(see also Yao et al. 2009). From our re-examination of the
fitting of the Ne edge in these data (obsid 1102) we find $N_{\rm H}$ = $(2.77\pm 0.72)\times 10^{21}$ 
atom cm$^{-2}$ (Table 2 and Fig. 7), in good agreement with the published value. It can be seen that 
the non-dip $N_{\rm H}$ value determined in the present work (Table 2) is higher but consistent with 
these values, (as is the non-dip value of $(3.0\pm 0.3)\times 10^{21}$ atom cm$^{-2}$ determined from
a reduced amount of data for Ne K). Moreover, the value obtained
from the Fe L edge (above) of $(2.5\pm 0.3)\times 10^{21}$ compares very well with
$(2.64^{+1.03}_{-0.34})\times 10^{21}$ atom cm$^{-2}$ obtained by Juett et al. (2006). Thus this also suggests
that no change has taken place. The difference between Ne and Fe values may indicate an underabundance of Fe,
i.e. a low Fe/Ne ratio as suggested by Juett et al. (2006). If we were to assume that the larger 
value in the present work from the Ne edge of $(3.40\pm 0.96)\times 10^{21}$
atom cm$^{-2}$ and the value of Juett et al. (2004) from the oxygen edge are actually different, the
implication could be that oxygen is depleted; however, Juett et al. (2004) found that the O/Ne ratio
in several sources was consistent with interstellar abundances; i.e. they found no evidence
for departure.

However, although the sizes of the uncertainties in column density in general 
indicate that there is no significant evidence that $N_{\rm H}$ has changed, one point,
that of obsid 1016, has a higher optical depth and a column density of $(4.04\pm 0.80)\times 10^{21}$ 
atom cm$^{-2}$, which is inconsistent with the point that follows it in Fig. 7. As the Galactic column
density cannot change, this provides some indication that there may be a low level of intrinsic absorption.
Evidence for intrinsic absorption was seen in Cygnus\th X-2 when in a much more active state 
by Ba\l uci\'nska-Church et al. (2010), $N_{\rm H}$ increasing between the soft and hard apex.
Evidence for a trend was clear, although the PCA instrument sensitive above 3 keV is not well
suited for accurate determinations of column density.

We also looked at the possibility of determining eventual changes in line emissions during the
persistent and dipping intervals. We detect all the lines described by Schulz et al. (2009) at line
fluxes that are reduced to an approximately similar fraction as the continuum flux. For the
selected persistent and dipping intervals the statistics are then already so low that we
could not engage in a meaningful differential analysis. An overall analysis of the
lines and a comparison with previous detections will be made in a future study.

\subsection{XMM OM results}

In Fig. 1 we compare the lightcurves of the Optical Monitor (lower panel) and the EPIC-pn (upper panel). 
Careful examination shows multiple occasions when X-ray dipping in the PN lightcurve is also seen in the 
OM lightcurve.
For example, at $\sim$42 ksec, there are definite flux decreases in both bands at 38.6, 40.4, 43.3 and 
45.7 ksec. Decreases are seen 
in both bands at 61.2 and 62.0 ksec and similar correspondences can be seen across the observation. The strong 
decreases in UV at 3 ksec and 30 ksec are matched by only weak decreases in the PN, suggesting that other factors
affect the UV emission. At times later than $\sim$50 ksec, the X-ray intensity drifts downwards
while the OM lightcurve drifts upwards, and attempts to derive formally a correlation coefficient
between OM and X-ray data are affected by this. However, this is the first time
that X-ray dipping in a LMXB has been seen in an optical or UV lightcurve.
The reduction of UV intensity during X-ray dips shows that a substantial part of the 
emission originates in the accretion disk and it thus appears that the absorbing material removes both 
X-ray and UV emission.
\begin{figure}[!h]                                                            % Fig. 8
\begin{center}
\includegraphics[width=76mm,height=76mm,angle=270]{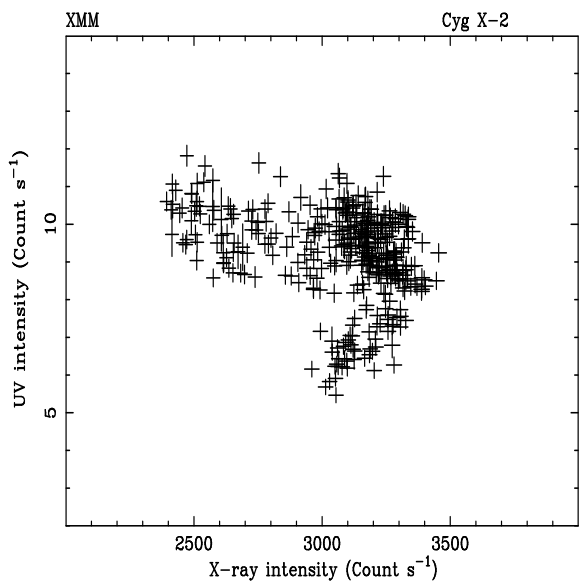}                 % was fig7_om_0.2-10.0_200s
\caption{Variation of Optical Monitor intensity in the UV band 2\th 500 - 4\th 000 \AA$\,$ 
with the X-ray intensity in the EPIC-pn 0.2 - 10.0 keV band. 
}
\label{}
\end{center}
\end{figure}

In Fig. 8 we show the variation of OM intensity with the EPIC-pn 0.2 - 10 keV intensity,
although this is confused by the drifts mentioned above. The X-ray dipping from 3400 to 2500
count s$^{-1}$ results in OM decreases from 10 to 9 count s$^{-1}$ forming data with a slight
slope in the figure. However, the drifts in both optical and X-ray intensities broaden this
branch with increased OM intensity at lower X-ray intensities, masking the correlation. The
other effect seen at 30 ksec of a deeper optical decrease for a smaller X-ray dip forms the 
second branch with a clear correlation. There may be an intrinsic variation in source intensity 
at this time so that a dependence of the optical signal on the X-ray intensity would indicate 
reprocessing of the X-ray emission. In any event, the small increase in the hardness ratios in Fig. 2 
at this time shows that weak X-ray dipping takes place and the existence of two branches in 
Fig. 8 suggests that two different processes are occurring.

\subsection{Search for QPO}

The EPIC-pn burst mode data were used to make a search for QPO. To avoid intoducing aliases,
the data were rebinned to 3.90625 ms, an exact multiple of the read-out time.
Barycentric corrected data were given standard filtering (Sect. 2.1) and lightcurves were extracted 
using {\sc xmmselect} and converted to power spectra using {\sc powspec}. This gave an accessible
frequency range of $10^{-1}$ to 128 Hz. The possible presence of kHz QPO could not be investigated 
because of the limits imposed by the binning required. The data were analysed in several ways.
Firstly, the complete observation was searched for QPO, divided into 16-second segments. The observation 
was also divided into 4096-second sections of data in each of which an average power spectrum
was extracted from shorter segments of data. Above 20 Hz a number of spurious peaks of instrumental
origin exist derived from the read-out time and other binnings. No significant QPO detections were found 
in the power spectra, although there was a possible weak peak at $\sim$ 5 Hz.

%A peak at a frequency of 0.684$\pm$0.023 Hz 
% 0.66 - 0.707
%was clearly detected with an amplitude of {\bf ??}. This appears similar to peaks at about 1 Hz 
%discovered in the dipping sources XB\th 1323-619, XB\th 1746-371 and XBT\th 0748-676
%(Jonker et al. 1999a,b and Homan et al. 1999). It was proposed that this was a new type of QPO 
%caused by quasi-periodic obscuration of the X-ray source by structure in the accretion disk, seen 
%when the source inclination is high. We report the first detection of a 1 Hz QPO in Cygnus\th X-2.

\section{Discussion}

\subsection{Position on the Z-track}

During the observations the source did not move appreciably along the Z-track, as can be seen
from the lack of intrinsic variation of the X-ray intensity in the EPIC-pn lightcurve (Fig. 1),
confirmed by the lack of movement in the hardness-intensity diagram (Fig. 2).
The only variability is due to the dipping and the results clearly show that the dips are due
to partial absorption events and are not source changes. Moreover, the lightcurve exhibits a lack of flaring, 
which would be seen as intensity increases lasting a few thousand seconds so that the source is not 
on the flaring branch. Indeed, in many observations Cyg\th X-2 does not exhibit a traditional FB 
in the hardness-intensity diagram, although this is occasionally seen. In addition there can be
intensity drops approximately at the soft apex, which form a distinct extra track in hardness-intensity. 
In the present observation, the neutron star 
blackbody temperature is low at $\sim$1.3 keV and previous work has shown that such low values are 
found in the Cygnus\th X-2 like Z-sources at or near the soft apex, $kT_{\rm BB}$ increasing to 
$\sim$2 keV at the hard apex and higher on the HB (e.g. Ba\l uci\'nska-Church et al. 2010). 
Moreover, we measure a blackbody radius $R_{\rm BB}$ of 9.6$\pm$ 1.1 km, indicating that the whole neutron star
is emitting, again as found in previous work to occur at the soft apex. It thus 
appears that the source remains in a fixed position on the Z-track in the neighbourhood of the soft 
apex and that dipping is seen but not flaring. Previous studies suggested that dipping was a characteristic 
of the flaring branch (Sect. 1; Hasinger et al 1990; Kuulkers et al. 1996), however, we now show 
that dipping and flaring are distinctly different phenomena. In terms of the model proposed for the
Z-track sources based on the extended ADC, the low value of $kT_{\rm BB}$ suggests a low and stable
mass accretion rate $\dot M$; an increase would drive the source towards the hard apex
and higher $kT_{\rm BB}$. It appears that $\dot M$ is indeed low, as the luminosity of the source 
is $\sim$30\% less than in the previous observations of 2007 (Schulz et al. 2009). It is apparent
that dipping is not a feature of the FB and furthermore, if flaring is due to unstable nuclear burning as 
proposed in the extended ADC model, and not due to $\dot M$ increase, there is no motivation for the 
suggestion that dipping is caused by inflation of the inner disk by radiation pressure at increased 
$\dot M$ (Kuulkers \& van der Klis 1995). 

The distinction between dipping and flaring is important to a fundamental question relating to the
Z-track sources. In a previous multi-wavelength campaign (Hasinger et al. 1990), Cygnus\th X-2 was observed
in X-ray, ultraviolet, optical and radio. The ultraviolet results (Vrtilek et al. 1990) showed an increase
of intensity on the FB and was a major argument for the widely accepted view that the mass accretion rate
increases on the Z-track monotonically in the sense HB - NB - FB. However, the hardness-intensity diagram
(Fig. 2 of Hasinger et al. 1990) viewed in the light of the present results clearly shows that the
source variation was {\it not flaring}, but dipping. Thus the major argument for $\dot M$ increasing 
on the FB disappears, as
does the contention that $\dot M$ increases from HB - NB - FB. In the Z-track source model based on an
extended ADC (above) it is argued that $\dot M$ does not change in flaring but increases between the soft
and hard apex, in the opposite sense to that in the standard view.

The lack of detection of radio emission is entirely consistent with this position on the Z-track: radio
detection from relativistic jets is known to be concentrated at around the hard apex, and little
emission is expected at the soft apex. In fact, it seems more plausible to us that any weak detection of
radio on the lower normal branch may be a residual from the source having previously been at the hard apex.
Spectral analysis of the Cyg-like sources as a function of Z-track position has previously shown
that the X-ray flux in the neighbourhood of the neutron star becomes several times the Eddington
flux. The strong correlation between this flux ratio $f/f_{\rm Edd}$ and radio detection thus
suggested that radiation pressure is responsible for jet launching via disruption of the
inner disk (Church et al. 2006).

\subsection{The nature of intensity dips}

The lack of intrinsic spectral variation of the source during our observation is rather fortunate because it 
allows the study of the dipping events without interfering spectral continuum changes. Furthermore, we 
can take advantage of the broadband properties of the {\it XMM} EPIC-pn and the high-resolution {\it Chandra} 
and {\it XMM} gratings in specific bandpasses. The EPIC-pn lightcurve shows dipping to remove only a fraction 
of the total luminosity ($\sim$25\%) and we can immediately infer that the absorber covers the X-ray source 
only partially, which in turn must also be extended. The results of spectral fitting to the EPIC-pn data 
demonstrate for the first time that dipping in Cygnus\th X-2 is due to very strong partial absorption. 
%The non-dip broadband column density agrees well with previous determinations (Juett et al. 2004); 
The spectrum is well-fitted by blackbody emission plus Comptonized emission, which we associate with the neutron 
star and the extended ADC, respectively. In addition, broad line emission is seen at $\sim$1 keV 
and at Mg XII, Si XIV, S XVI, and Fe XXV. The precise origin of the 1 keV line is still unknown and part 
of ongoing analysis.
The Comptonized emission dominates the X-ray flux, comprising 68\% of the flux in the band 
0.5 - 10 keV. The intermediate and deep dip spectra are well-fitted by a model in which the extended
Comptonized emission of the ADC is progressively overlapped by the absorber, and the covering factor rising 
to 0.4 in deep dipping with an associated column density of $4.5\times 10^{23}$ atom cm$^{-2}$. 
In fact, fitting does not permit absorption of the blackbody emission and this is remarkable, given 
that in the dipping class of LMXB in which dips are caused by absorption in the outer disk bulge, 
the neutron star blackbody emission is {\it always} removed (e.g. Church et al. 1997).
The dominant Comptonized emission in the present observation is removed gradually as dipping proceeds, 
i.e. it is partially covered with the covering factor increasing. This provides further strong support 
for the extended nature of the ADC in Cygnus\th X-2: part of the emitter being uncovered by absorber is 
not possible without an extended source.
%The partial nature of the absorber but also
%its strength and rapid variability argues against a simple increase of column density with a localized emitter.
In the dipping LMXB such as XB\th 1916-053 (Church et al. 1997), the same process of progressive
covering of an extended ADC also takes place, which is generally accepted as being due to absorption in the bulge in 
the outer accretion disk. However,
the suggestion that the dipping LMXB sources can be explained by an ionized absorber at some location closer 
to the neutron star, removing point-like Comptonized emission (Boirin et al. 2005, D\'iaz Trigo et al. 2006)
must be viewed with caution given the now extensive evidence for the extended ADC.

The {\it Chandra} gratings were used to provide high-quality spectra in the range
13 - 24 \AA $\,$ for both non-dip and deep dip data, revealing the Fe L~II (17.2~\AA) 
and L~III (17.5~\AA) edges, the O K edge (22.89~\AA)
and the Ne K edge at 14.29~\AA.  In the {\it Chandra} data the Ne edge has the best contrast
with respect to the continuum and an effort was made to compare with
results from previous observations. Most remarkable is the result that no difference in 
optical depth can be found between
non-dip and deep dip spectra. The optical depth in the deep-dip spectra is 0.097$\pm$ 0.025,
which is practically identical with the non-dip value of 0.108$\pm $0.030. The result is 
confirmed for the case of the O K edge in the RGS spectra. Given the large decrease in
continuum intensity during the dips and the associated large value of partial $N_{\rm H}$ found from the EPIC-pn,
a substantial increase in optical depth might have been expected. The fact that this is not found places
severe constraints on a physical explanation.

We have derived the orbital phase range of the present observations using the ephemeris of Casares 
et al. (2010) with an orbital period of 9.84450 days to be 0.324$\pm$0.007 - 0.425$\pm$0.007. 
This phase range is well outside the expected phase position of 0.75 where the disk bulge 
intercepts the line-of-sight between the X-ray emitting regions and the observer. We can  
thus rule out the bulge as the absorber, however, the actual phase of $\sim$0.35 agrees with a position
on the opposite side of the disk to the bulge where the accretion flow from the companion impacts.
Interdipping is seen in dipping sources such as XB\th 1916-053 with smaller dips between the main dips 
repeating at the orbital cycle (Church et al. 1997), caused possibly by ellipticity in the disk or by 
material overshooting the disk creating a secondary bulge. In the present case, we may be seeing
interdipping, or possibly we may see dipping caused by clouds of material traversing the system above the 
accretion disk. The inclination of the Cygnus\th X-2 system is known to be 62.5$\degmark$
(Cowley et al. 1979), a little outside the range of inclination angles of 65 - 85$\degmark$ normally 
associated with the dipping LMXB. At this inclination, we would expect an extended ADC to be partly
overlapped by absorber either at the main bulge or secondary bulge, but that the vertical height of
absorber would just miss overlapping the neutron star, thus explaining the lack of absorption of
the blackbody emission. Clouds of absorber traversing above the disk could also cover part of the ADC  
without covering the neutron star. Dipping appears entirely as absorption of the Comptonized emission. 
The fraction of this emission covered by absorber is strongly absorbed with a high column density 
$N_{\rm H}$ $\sim 4.5\times 10^{23}$ atom cm$^{-2}$ removing all flux below 4 keV. Thus, in the grating 
spectra a reduction of continuum intensity is seen in dipping, but we only see the 60\% of the emission 
that is not covered at all, and this explains why there is no increase in the measured optical depth in the edges.

We can estimate the size of absorber from dip ingress times. If dipping is caused by absorption in a 
secondary bulge in the outer accretion disk, the dip ingress time $\Delta t$ is the time needed for the absorber of
diameter $D$ and velocity $v$ to overlap the extended emitter, so that $\Delta t$ = $D/v$. The velocity 
at which the bulge moves
is $2\, \pi \, R/P$, where $P$ is the orbital period, and for the measured $\Delta t$ of $\sim$750 s,
the absorber size $D$ is 35\th 000 km. Schulz et al. (2009) give a radial size of the ADC in Cyg\th X-2
of 110\th 000 km, so that the absorber would convincingly give a covering factor of about 40\%, as observed.
%However, using the dependence of radial extent of the ADC on source luminosity in LMXB (Church \& 
%Ba\l uci\'nska-Church 2004), gives $R_{\rm ADC}$ $\sim$500\th 000 km, i.e. several times larger so that
%the agreement is not so good.  

If the absorbers were clouds traversing above the disk, these may be blobs of material originating where
the stream from the companion strikes the outer disk forming the bulge. Overflowing of the disk has been
suggested by several authors. Armitage \& Livio (1996) investigated theoretically the disturbance to the
accretion flow at the point of impact resulting in material being moved significantly out of the orbital 
plane although this was thought to return to the orbital plane and not cross the disk to the opposite side.
Assuming a typical infall velocity that will be much higher than the bulge velocity
(above) would lead to a very large absorber size $D$ = $v\, \Delta t$, several times larger than 
for the bulge (above), so giving 100\% dips, much deeper than we observe.

%The ingress time is again $\Delta$t = D/v where $v$ is the radial infall velocity of a cloud at a radius 
%equal to that of the ADC. Using a radius of 110\th 000 km (above) gives an infall velocity of $\sim$2000
%km s$^{-1}$ and a size $D$ of $1.4\times 10^6$ km. This absorber size is larger than the 
%assumed ADC sizes, so dipping would be 100\% deep not as observed, indicating that this model is unlikely 
%and that absorption takes place in the outer disk. }

\subsection{Is there a dipping state ?}

Last, but not least, we would like to address the issue of why we observe these dipping events in
Cygnus\th X-2 when the source is near the soft apex. In the previous {\it Chandra} observation
the dipping occurred at the soft apex (Schulz et al. 2009), in the present one the derived blackbody
temperature and radius imply a similar location on the Z-track, albeit at an overall lower flux.
The persistent X-ray flux is dropping at a constant rate indicating the source may still be moving to 
a lower flux. At the soft apex, the spectra appear particularly soft and the X-ray intensity is low.

The reason why the overall intensity in our observation is much lower is not clear although it
may well be due to reduced mass accretion rate. We can, however,  
exclude enhanced absorption. Figure 7 compares the optical depths of the Ne K edge for previous 
high-resolution observations and as discussed previously, there is no significant increase in
the present observation. The values obtained in non-dip from the three instruments: 
($3.6\pm 0.1)\times 10^{21}$ (EPIC-pn), ($3.57^{+0.14}_{-0.17})\times 10^{21}$ (RGS) and 
$(3.4\pm 0.96)\times 10^{21}$({\it Chandra}) atom cm$^{-2}$ agree well. If we were to assume that 
the difference between these values and the Juett et al. (2004) value of $(2.3\pm 0.5)\times 10^{21}$ 
atom cm$^{-2}$ was real, we find from our EPIC-pn spectral fit to non-dip emission that increasing $N_{\rm H}$ 
from the Juett value to the EPIC value causes the 1 - 10 keV luminosity to decrease by only 5\%
so it is ruled out that the decrease in luminosity since the previous {\it Chandra} observation is 
due to enhanced absorption.

%{\bf It is clear that we do not see dipping on the flaring branch, since this is rarely seen. It is not clear
%why we apparently do not see dipping on the HB. On the hardness-intensity diagram, dipping may cause
%motion along the existing HB and so not be visible; in the lightcurves, the timescale for dipping can 
%be similar to that for motion along the track making dipping less obvious.

The fact that we often observe dipping at the soft apex might simply indicate that we observe a switch 
from a mostly ionized plasma on the normal branch and horizontal branch to the regaining of ionization 
fractions at lower luminosities. At the soft apex the plasma begins to recombine even to the point of
neutral material.
%
%Observing dipping at the soft apex may suggest some modification of the absorber at high source
%luminosities. For an absorber in the outer disk as we think, it is difficult to see how an increase
%in luminosity of less than 50\% as the source moves away from the apex would result in a total cessation 
%of absorption in the outer disk. For the less likely case of absorption in clouds above the disk, the 
%material may be fully ionized on the NB and HB, but at the soft apex begin to recombine and neutral matter appear. 
For recombination coefficients of $\sim 10^{-11}$ cm$^3$ s$^{-1}$ and 
plasma densities above 10$^{12}$ cm$^{-3}$ this should happen in a fraction of a second. However, 
detailed modelling would be required to fully understand the ionization balance in such a scenario. 
In summary, it remains unclear why dipping appears concentrated at this part of the Z-track.

%seems that the source in the hardness-intensity diagram is resting at 
%the soft apex while the accretion rate is low. This might not constitute
%a particular new source state but it remains to be explained why the source is
%not moving either back into the normal or is not proceeding onto the flaring branch.

\thanks{
We thank the participants in the e-EVN radio observations at Cambridge, Jodrell Bank, Knockin, Effelsberg,
Medicina, Torun, Onsala, Sheshan and Yebes. This work was supported in part by the Polish Ministry of 
Science and Higher Education grant 3946/B/H03/2008/34.}

\end{document}